\documentclass[aps,pra,onecolumn,superscriptaddress,10pt]{revtex4}

\usepackage{latexsym}
\usepackage{amsfonts}
\usepackage{amsmath}
\usepackage{graphicx}
\usepackage{dcolumn}
\usepackage{url}

\usepackage{setspace}
\usepackage{natbib}

\usepackage{upgreek}

\providecommand\bnabla{\boldsymbol{\nabla}}
\providecommand\bcdot{\boldsymbol{\cdot}}

\newsavebox{\astrutbox}
\sbox{\astrutbox}{\rule[-5pt]{0pt}{20pt}}

\newcommand{\rmd}{{\mathrm{d}}}
\newcommand{\rme}{{\mathrm{e}}}
\newcommand{\rmi}{{\mathrm{i}}}

\newcommand{\lamD}{{\lambda_{\mathrm{D}}}}
\newcommand{\lma}{{\lambda_{\mathrm{ma}}}}

\newcommand{\bsa}{{\boldsymbol{a}}}
\newcommand{\bsb}{{\boldsymbol{b}}}
\newcommand{\bsk}{{\boldsymbol{k}}}
\newcommand{\bsm}{{\boldsymbol{m}}}

\newcommand{\bsr}{{\boldsymbol{r}}}
\newcommand{\bsv}{{\boldsymbol{v}}}

\newcommand{\itPhi}{{\mit \Phi}}

\newcommand{\upi}{{\uppi}}

\renewcommand{\S}{Sec.\  }

\newcommand\norm[1]{{\| #1 \|}}

\begin{document}

\def\Marseille{UMR 7345 CNRS, Aix-Marseille Universit\'e, 
campus Saint-J\'er\^ome, case 321, \\ 
av.\ esc.\  Normandie-Niemen, FR-13397 Marseille cedex 20}

\title{Uniform derivation of Coulomb collisional transport thanks to Debye shielding}

\author{ D~F Escande, Yves Elskens and F Doveil}
\affiliation{\Marseille}
\email{dominique.escande@univ-amu.fr, yves.elskens@univ-amu.fr, fabrice.doveil@univ-amu.fr}







\begin{abstract}
The effective potential acting on particles in plasmas being essentially the Debye-shielded Coulomb potential,
the particles collisional transport in thermal equilibrium is calculated for all impact parameters $b$,
with a convergent expression reducing to Rutherford scattering for small $b$.
No cutoff at the Debye length scale is needed, and the Coulomb logarithm is only slightly modified.
%
%
%
\newline 
{\bf{PACS numbers}}~:  \newline
      {52.20.-j}  {Elementary processes in plasmas}  \newline 
      {45.50.-j}  {Dynamics and kinematics of a particle and a system of particles}   \newline 
      {52.25.Fi}  {Transport properties}  \newline 
       {52.25.Dg}  {Plasma kinetic equations}

\end{abstract}


\maketitle



\section{What do we call collisions and what is their effect~?}
\label{sec:Intro}

Transport properties of plasmas are challenging in many ways.
Because of the long range nature of the Coulomb interaction,
a first large-scale description of the plasma is given by the Vlasov equation,
which is derived in the mean-field limit directly from the many-body equations of motion
under the condition that the interaction be regular at short range
\citep[see][for a clear, short and rigorous derivation]{Spohn91}.
However, the Coulomb potential is too singular at short range to allow for such a derivation of the Vlasov equation
\citep[see][for a review and alternative approaches]{Kie14,EED_Vlasovia}.
As the Vlasov limit neglects the contribution of two-body correlations to the total force acting on a particle
in the $N \to \infty$ limit, where this force is lumped into a smooth mean field,
the obstacle imposed by the Coulomb interaction to this limiting description is summarized under
the heading ``collisions''.

Such a heading is a result of the development of plasma physics after that of gases,
which made natural for the former to borrow concepts and tools from the latter.
In particular, the unscreened interactions of particles in kinetic plasmas were considered as collisions.
Indeed, these interactions induce modifications to the Vlasov equation germane to the Boltzmann collision operator
\citep[see][for a classical treatment, with plasmas briefly discussed in ch.~14]{FerKa72}.
It must be noted, however,
that the mathematical understanding of collisions in gases, from the microscopic reversible newtonian mechanics,
was listed by Hilbert among important problems for the XXth century,
and it remains a major research issue,
on which Saint-Raymond and coworkers \citep{BoGaStRay14,GalStRayTex} just accomplished two breakthroughs.
For plasmas, progress in the understanding of reversible roots of macroscopic irreversibility
has been made for the quasilinear description of wave--particle systems
\citep{BE97, BE98, EEbook, BEEB, ElPa10, El12},
where momentum exchange between particles and waves may be viewed like a collision,
for which ``local interactions'' take place in velocity space rather than in position space.

However, the interaction of particles in a plasma corresponds seldom to two-body collisions,
even when taking into account Debye shielding~:
in a plasma where the interparticle distance is much smaller than the Debye length,
a particle $j$ feels the simultaneous unscreened short-range action of many particles.
Except for those particles very close to $j$, this action produces a slow and simultaneous deflection of $j$.
Rigorously speaking, one should not speak about collisions, but about ``short range induced interactions",
``unscreened Coulomb interactions", or so.
For simplicity and to stick to the traditional vocabulary in textbooks, we use ``collisions'' in the following.

Almost sixty years ago, two groups at UC Berkeley's Radiation Laboratory
simultaneously studied transport due to collisions in non-magnetized plasmas,
and they quoted each other's results in their respective papers~:
one in 1956 by Gasiorowicz, Neuman and Riddell and,
in 1957, one by Rosenbluth, MacDonald and Judd.
The second group of authors used the Rutherford picture of two-body collisions, while
the first group of authors dealt with the mean-field part of the interaction by using perturbation theory in electric field amplitude.
Within the same approximations as \citet{Gasio},
a more elegant derivation of the collisional transport coefficients was provided,
in a ``post-vlasovian'' approach,
by taking the limit ``infinite number of particles in the Debye sphere" of the Balescu--Lenard equation
(see section 8.4 of \citet{BalBk97} and sections 7.3 and 7.4 of \citet{HaWa04}),
though the rigorous foundation of this equation is still a challenge \citep{Spohn91,Lancellotti10}.
With a single calculation, the Balescu--Lenard approach provides both the dynamical friction and the diffusion coefficient.

For the following discussion, it is useful to introduce the characteristic lengths~:
(i) the interparticle distance $d = n^{-1/3}$ where $n$ is the plasma density~;
(ii) the classical distance of minimum approach $\lambda_{\rm{cma}} = e^2 /(4 \upi \epsilon_0 k_{\rm{B}} T)$
where $\epsilon_0$ is the vacuum permittivity, $k_{\rm{B}}$ is the Boltzmann constant, $T$ is the temperature,
and $e$ is the electron charge~;
(iii) the Debye length $\lamD = [(\epsilon_0 k_{\rm{B}} T)/(n e^2)]^{1/2} = d^{3/2} / (4 \upi \lambda_{\rm{cma}})^{1/2}$.
Recall that $\lambda_{\rm{cma}} \ll d \ll \lamD$
in a plasma with a large number of particles in the Debye sphere.

Now, we can point out that each of the above works on collisional transport
has a difficulty in describing the interactions at distances of the order of the typical interparticle distance $d$.
Indeed, the mean-field approach cannot describe the graininess of these scales,
and the Rutherford picture cannot describe the simultaneous collisions with several particles.
Consequently, the mean-field approach is suited to describing scales larger than $d$,
and should be used with a corresponding ultraviolet cutoff,
while the Rutherford picture holds for scales smaller than $d$, and should be used with a corresponding infrared cutoff.
Fortunately, in both approaches the transport coefficients depend only logarithmically on these cutoffs.
Furthermore, forgetting about the latter ones, and considering in both cases the scales
typically between $\lambda_{\rm{cma}}$ and $\lamD$, the two results are found to agree \citep{Gasio,RoMcDJu}.
This provided confidence in these complementary extrapolations which are the present basis
of the description of collisional transport in plasmas, as presented in many plasma physics textbooks.

However, as yet a calculation of the contribution of scales about $d$ to collisional transport has been missing,
and no theory provides a calculation of this transport covering all scales between $\lambda_{\rm{cma}}$ and $\lamD$.
It is this gap that the present paper aims at filling.
The basic idea of the new derivation is to substitute the bare Coulomb potential of a particle
with its ``dressed'' Debye-shielded potential.
Kinetic theory is traditionally used to introduce this shielded potential \citep{Gasio,RostRos60,BalBk63,Rost64},
but this can also be done by using a direct, perturbative analysis of the particles motion
in an $N$-body description of the plasma without needing to introduce any test particle \citep{EDE13}.
Here, by using the shielded potential in the plasma,
the trace $T_D$ of the velocity diffusion tensor of a given particle is computed by a convergent expression
including the particle deflections for all impact parameters.
These deflections are computed by first order perturbation theory in the total electric field,
except for those due to close encounters.
The contribution to $T_D$ of the former ones is matched with that of the latter ones computed by \citet{RoMcDJu}.
The detailed matching procedure includes the scale of the inter-particle distance,
and is reminiscent of that of \citet{Hub}, however without invoking the cancellation of three infinite integrals.
It leads to the same expression as \citet{RoMcDJu},
except for the Coulomb logarithm which is modified by a velocity dependent quantity of the order of 1.
The structure of the derivation is outlined just before \S\ref{sec:PerturbTraj}.

\section{Collisional transport in plasmas}
\label{sec:CollTrans}

Consider a plasma in thermal equilibrium, with a uniform density,
in which the particles have random initial positions.
Then the dynamics of particles has no collective aspect, but is ruled by the cumulative effect of two-body deflections.
More specifically, we choose random initial positions $\bsr_{l0}$'s and assign to each particle a well defined velocity $\bsv_{l0}$,
in such a way that the overall initial smooth velocity distribution is close to some given function.
To formulate the dynamics as a finite-dimensional system of differential equations,
we consider $N$ electrons in a cube with size $L$, with periodic boundary conditions
(this is equivalent to immersing the electrons in a uniform neutralizing background),
and let $L \to \infty$, $N \to \infty$ with constant particle density $n = N/L^3$ (and hence constant Debye length $\lamD$).

The first effect of Coulomb repulsion between the electrons is to slightly alter their motions,
$\bsr_l (t) = \bsr_{l0} + \bsv_{l0} t + \delta \bsr_l (t)$,
in such a way that their interaction is well described by the shielded Coulomb interactions \citep{EDE13},
i.e.\ we write
\begin{equation}
  \delta \ddot{\bsr}_l
  =
  \sum_{j \in S ; j \neq l} \bsa(\bsr_l-\bsr_j,\bsv_j),
\label{delrsecscreen}
\end{equation}
with $S$ denoting the set of integers from 1 to $N$ labeling particles, and
\begin{equation}
  \bsa(\bsr,\bsv)
  =
  \frac{e}{m_\rme} \bnabla \itPhi (\bsr,\bsv),
\label{acc}
\end{equation}
with $-e$ and $m_\rme$ the electron charge and mass, and with the effective potential
\begin{equation}
  \itPhi (\bsr,\bsv)
  =
  - \frac{e}{L^3 \epsilon_0} \sum_{\bsm} \frac{\exp(\rmi \bsk_{\bsm} \bcdot \bsr)}
                                                                      {k_{\bsm}^2 \epsilon(\bsm,\bsk_{\bsm} \bcdot \bsv)} .
  \label{phi}
\end{equation}
Here, 
$\epsilon(\bsk_\bsm, \omega)$ is the dielectric function of the plasma.
The space Fourier transform is defined with the wave vectors
$\bsk_{\bsm} = 2 \upi \, \bsm / L$ (and $k_{\bsm} = \|\bsk_{\bsm}\|$), where the sum runs over all vectors
$\bsm = (m_x,m_y,m_z)$ with three integer components.
For simplicity, we focus below on slow particles, so that
$\epsilon(\bsm,\mathbf{0})  = 1 + (k_{\bsm} \lamD)^{-2}$
and $\itPhi$ reduces to the Yukawa potential with decay length $\lamD$.

We compute the deflection of particle $l$ in a sequence of steps.
In \S\ref{sec:PerturbTraj}, we use first order perturbation theory in $\itPhi$,
which shows the total deflection to be the sum of the individual deflections due to all other particles.
For an impact parameter much smaller than $\lamD$,
the deflection due to a particle reduces to the Rutherford deflection due to this particle as if it were alone.
In \S\ref{sec:CloseColl}, for a close encounter with particle $j$, we show that
the deflection of particle $l$ is exactly the one it would undergo if the other $N-2$ particles were absent.
In \S\ref{sec:SmallDeflec}, we show that the deflection for an impact parameter of the order of $\lamD$ is given by
the Rutherford expression multiplied by some function of the impact parameter reflecting shielding.
Finally, since the individual deflections due to impact parameters $b$ exceeding $\lamD$ decay rapidly with $b$,
these three steps yield an analytical expression for deflection whatever the impact parameter,
as discussed in \S\ref{sec:AllColl}.

\subsection{Perturbative approximation to trajectories}
\label{sec:PerturbTraj}

We first compute $\delta \bsr_l$ by first order perturbation theory in $\itPhi$,
taking the ballistic motion $\bsr_l^{(0)} (t) = \bsr_{l0} + \bsv_{l0} t$ as zeroth order approximation.
This yields
\begin{equation}
  \delta \dot{\bsr}_l(t)
  =
  \sum_{j \in S ; j \neq l} \delta \dot{\bsr}_{lj}(0,t) ,
\label{delrl1}
\end{equation}
where
\begin{equation}
  \delta \dot{\bsr}_{lj}(t_1,t_2)
  =
  \int_{t_1}^{t_2}  \bsa[\bsr_l^{(0)}(t')-\bsr_j^{(0)}(t'),\bsv_j] \, \rmd t'.
\label{delrlj1}
\end{equation}
It is convenient to write
\begin{equation}
  \bsr_l^{(0)}(t')-\bsr_j^{(0)}(t')
  =
  \bsb_{lj}+ \Delta \bsv_{lj} (t' - t_{lj}),
  \label{blj}
\end{equation}
where $t_{lj}$ is the time of closest approach of the two ballistic orbits,
and $\bsb_{lj}$ is the vector joining particle $j$ to particle $l$ at this time.
Then $b_{lj} = \| \bsb_{lj}\|$ is the impact parameter of these two orbits when singled out.
The initial random positions of the particles translate into random values for $\bsb_{lj}$ and $t_{lj}$.
The typical duration of the deflection of particle $l$ given by Eq.~(\ref{delrlj1}) is $\Delta t_{lj} \equiv b_{lj} / \Delta v_{lj}$
where $\Delta v_{lj} = \| \Delta \bsv_{lj} \|$,
but a certain number, say $\alpha$, of $\Delta t_{lj}$'s are necessary for the deflection to be mostly completed.
For a given $b_{lj} $ and for $t \gg \Delta t_{lj}$ in Eq.\ (\ref{delrl1}),
the deflection of particle $l$ given by Eq.~(\ref{delrlj1}) is maximum
if $t_{lj}$ is in the interval $[\alpha \Delta t_{lj}, t - \alpha \Delta t_{lj}]$.
We notice that $\Delta t_{lj}$ is about the inverse of the plasma frequency for $b_{lj} \sim \lamD$
and $\Delta v_{lj}$ on the order of the thermal velocity.

For brevity, we compute here only the trace of the diffusion tensor for the particle velocities.
To this end, we perform an average over all the $\bsr_{l0}$'s to get
\begin{equation}
  \langle \norm{\delta \dot{\bsr}_l (t)}^2 \rangle
  =
  \sum_{j \in S ; j \neq l}  \langle \norm{\delta \dot{\bsr}_{lj} (t)}^2 \rangle,
\label{delrlj1^2}
\end{equation}
taking into account Eq.~(\ref{phi}), and the fact that the initial positions are independently random,
as well as the $\bsr_i - \bsr_j$'s for $i \neq j$.
Therefore, though being due to the simultaneous scattering of particle $l$ with the many particles inside its Debye sphere,
$\langle \norm{\delta \dot{\bsr}_l (t)}^2 \rangle$ turns out to be the sum of individual two-body deflections
for $b_{lj} $'s such that first order perturbation theory is correct.
\emph{Hence the contribution to $\langle \norm{\delta \dot{\bsr}_l (t)}^2 \rangle$ of particles with given $b_{lj}$ and $\Delta v_{lj}$
can be computed as if it would result from successive two-body collisions,
as was done by \citet{RoMcDJu} and in many textbooks}.

For $b_{lj} \ll \lamD$,
the main contribution of $\bsa[\bsr_l^{(0)}(t')-\bsr_j^{(0)}(t'),\bsv_j]$ to the deflection of particle $l$
comes from times $t'$ at which $\| \bsr_l^{(0)}(t')-\bsr_j^{(0)}(t') \| \ll \lamD$.
Therefore $\bsa(\bsr,\bsv)$ takes on its bare Coulombian value,
and $\delta \dot{\bsr}_l(t)$ is a first order approximation
of the effect on particle $l$ of a Rutherford collision with particle $j$.
Both the approximate value and the Rutherford collision value for $\delta \dot \bsr_{lj}$
scale like $\lma \Delta v_{lj} / b_{lj}$,
where $\lma = {e^2}/( \upi \, m_\rme \epsilon_0 \Delta v_{lj}^2)$ is
the distance of minimum approach of two electrons in a Rutherford collision, as allowed by energy conservation.
As the approximate value differs from the exact one by a factor
$O(\norm{\delta \dot \bsr_{lj}(-\infty,+\infty)}/\Delta v_{lj}) = O(\lma / b_{lj})$,
the perturbative calculation is seen to be correct for $b_{lj} \gg \lma$,
as long as the sum of deflections remains small compared with $\Delta v_{lj}$.

Summing over many collisions from $0$ to $t$ to estimate $\langle \norm{\delta \dot \bsr_l (t)}^2 \rangle$ in Eq.\ (\ref{delrlj1^2})
preserves the relative accuracy of the estimate for the contribution
of these intermediate range ($\lma \ll b_{lj} \ll \lamD$) deflections to the diffusion coefficient.
Note also that the small deflections $\delta \dot \bsr_{lj}$ are elastic,
which implies that they are orthogonal to the relative velocities $\Delta \bsv_{lj}$ to first order
(indeed they are parallel to their $\bsb_{lj}$'s to this order).
Higher order perturbation theory finds the projection of $\delta \dot \bsr_{lj}$ along $\Delta \bsv_{lj}$ to be of second order.

\subsection{Close collisions}
\label{sec:CloseColl}

Second, we consider the case of a close approach of particle $j$ to particle $l$, i.e.\ $b_{lj} \sim \lma$.
We write the acceleration of particle $l$ as
\begin{equation}
  \ddot{\bsr}_l
  =
  \bsa(\bsr_l-\bsr_j, \bsv_j) + \sum_{p \in S ; p \neq l,j} \bsa(\bsr_l-\bsr_p, \bsv_p).
  \label{delrsecEx}
\end{equation}
For particle $j$, we write the same equation by exchanging indices $l$ and $j$.
Since the two particles are at distances much smaller than the inter-particle distance $d = n^{-1/3} = N^{-1/3} L$,
the accelerations imparted to them by all other particles are almost equal.
Therefore, when subtracting the two complete equations of motion, the two summations over $p$ almost cancel, leaving
\begin{equation}
  \frac{\rmd^2 (\bsr_l - \bsr_j)}{\rmd t^2}
  = 2 \bsa(\bsr_l-\bsr_j),
  \label{delrsecEx2}
\end{equation}
which is the equation describing the Rutherford collision of these two particles in their centre-of-mass frame,
in the absence of all other particles (at such distances the shielded potential is the bare Coulomb one).
Since $b_{lj} \ll d$, $\Delta t_{lj}$ is much smaller than the $\Delta t_{lp}$'s of the other particles.
Therefore the latter produce a negligible deflection of the centre of mass during the Rutherford two-body collision,
and the deflection of particle $l$ during this collision is exactly that of a Rutherford two-body collision.
The contribution of such collisions to $\langle \norm{\delta \dot \bsr_l (t)}^2 \rangle $ was calculated by \citet{RoMcDJu}.

Now, since the deflection of particle $l$ due to particle $j$ as computed by the perturbation theory of \S\ref{sec:PerturbTraj}
is an approximation to the Rutherford deflection for the same impact parameter,
we may conversely approximate the perturbative deflection with the full Rutherford one,
and obtain an obvious matching of the theories
for $b_{lj} \sim \lma$ and for $\lamD \gg b_{lj} \gg \lma$~:
we may thus use the estimate of \citet{RoMcDJu} in the whole domain $b_{lj} \ll \lamD$.

\subsection{Small deflections}
\label{sec:SmallDeflec}

Third, we deal with impact parameters of the order of $\lamD$.
Then the deflection due to particle $j$ must be computed with Eq.~(\ref{delrlj1}).
For simplicity, we do the calculation for the case where $\bsv_j$ is small,
so that $\itPhi (\bsr,\bsv) \simeq \itPhi (\bsr,\mathbf{0})$
which is the Yukawa potential
$\itPhi_{\rm{Y}} (\bsr) = - e \, (4 \upi \, \epsilon_0 \| \bsr \|)^{-1} \exp (- \| \bsr \| / \lamD)$
\citep[Eq.~(18) of][]{Gasio}.
The first order correction in $\bsk_{\bsm}  \bcdot \bsv_j$ to this approximation
is a dipolar potential with an electric dipole moment proportional to $\bsv_j$.
Since a Maxwell distribution is an even function of $\bsv$, these individual dipolar contributions cancel globally.
As a result, the first relevant correction to the Yukawa potential is of second order in $\bsk_{\bsm}  \bcdot \bsv_j$.
This should make the Yukawa approximation relevant for a large part of the bulk of the Maxwell distribution.

In the small deflection limit, a standard calculation using the fact that the force derives from a central potential
shows the full deflection of particle $l$ due to particle $j$ to be
\begin{equation}
  \delta \dot{\bsr}_{lj}(- \infty,+ \infty)
  =
  \frac{e^2}{ 4 \upi m_\rme \epsilon_0} \, \bsb_{lj}
    \int_{- \infty}^{+ \infty}  \left[\frac{1}{r^3(t)} + \frac{1}{\lamD r^2(t)}\right] \exp [- \frac{r(t)}{\lamD}] \, \rmd t,
  \label{delrljT}
\end{equation}
where $r(t) = (b_{lj}^2 +\Delta v_{lj}^2 t^2)^{1/2}$ and $\bsb_{lj}$ was defined with Eq.~(\ref{blj}).
On introducing the angle $\theta = \arcsin [ \Delta v_{lj} t / r(t)]$, this integral becomes
\begin{equation}
  \delta \dot{\bsr}_{lj}(- \infty,+ \infty)
  =
  - \frac{2 e^2}{ 4 \upi m_\rme \epsilon_0 \Delta v_{lj}}
     \, \frac{h(b_{lj})}{b_{lj}^2 } \, \bsb_{lj},
\label{delrljTfin}
\end{equation}
where
\begin{equation}
  h(b)
  =
  \int_{0}^{\upi/2} \left[ \cos \theta +  \frac{b}{\lamD} \right] \exp [- \frac{b}{\lamD \cos \theta}] \, \rmd \theta
\label{delrljT2}
\end{equation}
During time $t \gg \Delta t_{lj}$, a volume $2 \upi \, \Delta v_{lj}\, t\, b_{lj}\, \delta b_{lj}$ of particles
with velocity $\bsv_j$ and impact parameters between $b_{lj}$ and $b_{lj} + \delta b_{lj}$
produce the deflection of particle $l$ given by Eq.~(\ref{delrljTfin}),
and a contribution scaling like $[h^2(b_{lj})/b_{lj}] \delta b_{lj}$ to $\langle \norm{\delta \dot{\bsr}_l (t)}^2 \rangle$.

\subsection{Synthesis over all collision scales}
\label{sec:AllColl}

Let $b_{\rm{min}}$ be such that $\lamD \gg b_{\rm{min}} \gg \lma$.
The contribution of all impact parameters between $b_{\rm{min}}$ and some $b_{\rm{max}}$
is thus scaling like the integral $\int_{b_{\rm{min}}}^{b_{\rm{max}}} [h^2(b)/b]  \ \rmd b$.
Since $h(0) \simeq 1$ for $b$ small, if $b_{\rm{max}} \ll \lamD$,
this is the non-shielded contribution of orbits relevant to the above perturbative calculation.
Now recall that, on approximating it with the Rutherford-like result of \citet{RoMcDJu},
this contribution matches the contribution of impact parameters on the order of $\lma$.
Thus the contribution of all impact parameters
between $\lma$ and some $b_{\rm{max}}$ (small with respect to $\lamD$)
is scaling like the integral $\int_{\lma}^{b_{\rm{max}}} (1/b)  \ \rmd b$
as was computed by \citet{RoMcDJu}.
The matching of this result for $b_{\rm{max}} \gtrsim \lamD$ is simply accomplished
by keeping the factor $h^2(b)$ in the integrand,
which makes the integral converge for $b \to \infty$.
Taking this limit, one finds (see Appendix) that
the Coulomb logarithm $\ln (\lamD / \lma)$ of the second Eq.~(14) of \citet{RoMcDJu}
becomes $\ln (\lamD / \lma) + C$ where $C$ is of order unity.
If the full dependence of the shielding on $\bsv_j$ were taken into account \citep[see {e.g.}][]{Dewar,DewarL},
the modification to the Coulomb logarithm would be velocity dependent.

\section{Summary and perspectives}
\label{sec:Conclusion}

It is known that, in plasmas,
the Coulomb interaction spontaneously generates particle motions which alter the ``bare'' Coulomb pair interaction \citep{Gasio,RostRos60,BalBk63,Rost64,EDE13}.
In dilute, warm plasmas, particle trajectories are almost ballistic,
but any electron, say $j$, slightly affects all other electrons, so that $\bsr_p(t) = \bsr_{p0} + \bsv_{p0}t + \delta \bsr_p(t)$.
As a result, the total force on an electron $l$ due to electron $j$ is, to dominant order,
the sum of their direct Coulomb interaction $m_\rme \bsa_{\mathrm{C}}(\bsr_l-\bsr_j)$
(with $\bsa_\mathrm{C}(\bsr) = e^2 \bsr / (4 \upi \epsilon_0 m_\rme \norm{\bsr}^3)$)
and of the force on $l$ summing corrections to the Coulomb interaction with all other electrons
$- m_\rme \sum_{p \neq l,j} \nabla \bsa_\mathrm{C}(\bsr_j - \bsr_p) \bcdot \delta \bsr_p$ \citep{EDE13}.
The balance of these effects generates dynamically the Debye screening,
which thus results from the Coulomb interaction mediated by the plasma.
In a sense, Debye screening is the result of small deflections, which one is tempted to call collisions.

It is somewhat startling that, in turn, the resulting Debye screened effective potential
yields a description of pair interaction which provides a direct calculation of particle deflections,
viz.\ of collisional transport. Screening and collisions are thus intimately linked,
and our ability to calculate collisional transport rests on this link.

A second startling aspect of collisions in plasmas is that,
although each particle interacts simultaneously with many other ones on the Debye length scale
(suggesting the need for a collective description),
the transport effect of these interactions is well approximated by a sum of independent binary estimates,
because the deflections are so weak that they can be treated perturbatively.
This paradox may lead to misunderstandings in the description of the calculations.

The calculation of dynamical friction, which requires second order perturbation theory,
follows the same lines as those for the diffusion coefficient.
For the sake of simplicity, we computed here only the trace of the diffusion tensor~;
the same argument could be easily applied to the elements of the tensor.
It also extends to the tensors corresponding to electron-ion collisions and to ion-ion collisions.
For an inhomogeneous plasma, the acceleration of particle $l$ may be split into a homogeneous and a wave part,
so that the diffusion coefficient and the dynamical friction,
estimated by perturbative calculation of the dynamics up to second order,
are the sum of the collisional contribution and of a contribution due to waves,
the latter as calculated for instance in \citet{EZE96} and \citet{EEbook}.
We defer these issues to a later publication.

We computed here only the contribution to the trace of the diffusion tensor coming from particles slow enough for a Yukawa potential to be a good approximation for their shielded potential, which made possible an analytical estimate. 
The contribution of faster particle involves a more intricate shape of the shielded potential that does not look as analytically tractable \citep{Dewar,DewarL}, and will probably require subtle computer integration. 
This means a large amount of work which is out of the scope of our present paper. 



DFE acknowledges fruitful discussions with participants to the meeting
``Equilibrium and out-of-equilibrium properties of systems with long-range interactions'' at ENS-Lyon (August 2012).
YE enjoyed discussions with participants to Vlasovia in Nancy (November 2013).

\appendix
\section{Convergent integral for large impact parameter}
\label{appA}

The contribution of small deflections to the sum in Eq.\ (\ref{delrlj1^2}) is easily estimated.
Given a particle $l$, for a plasma in thermal equilibrium,
the relative velocity $\Delta v_{lj}$ and the impact parameter $\bsb_{lj}$
may be considered independent in first approximation.
The contribution of all particles with a velocity $\bsv_j$ to the velocity diffusion of particle $l$ is then
\begin{equation}
  \langle \norm{\delta \dot \bsr_l}^2 \rangle_{\Delta v_{lj}}
  =
  \left( \frac{e^2}{2 \upi \, m_\rme \epsilon_0 \Delta v_{lj}} \right)^2 2 \upi \, \Delta v_{lj}Êt \, I (\lma)
\label{A:estimD}
\end{equation}
where we define $I(\lambda) = \int_\lambda^\infty Êh^2(b) \, b^{-1} \, \rmd b$.
Replacing $h(b)$ with the step function $1$ for $0 < b < \lamD$ (and $0$ otherwise) yields the usual
\begin{equation}
  I_0(\lma)
  =
  \int_{\lma}^\infty Êh_0^2(b) / b \ \rmd b
  =
  \int_{\lma}^{\lamD} Ê1/ b \ \rmd b
  =
  \ln  \frac{\lamD}{\lma}
\label{logC}  .
\end{equation}
For the actual integral $h$ defined in Eq.\ (\ref{delrljT2}), the divergence for small $\lma$ is identical. Let us thus write
\begin{equation}
  I(\lma)
  =
  \int_{\lma}^\infty Êh^2(b) / b \ \rmd b
  =
  I_0(\lma) + I_1(\lma)
\label{I0I1}
\end{equation}
where we define for $0 < \lma  \leq \lamD$
\begin{equation}
  I_1(\lma)
  :=
  I_{11} + I_{12}
  =
  \int_{\lma}^{\lamD}Ê\frac{h^2(b) - 1} {b} \rmd b
  +
  \int_{\lamD}^\inftyÊ\frac{h^2(b)} {b} \rmd b   .
  \label{I1sum}
\end{equation}
Note that $I_{11} < 0$ and $I_{12} >0$. We now estimate both integrals.

The upper estimate (using $0 \leq \cos \theta \leq 1$ in Eq.\ (\ref{delrljT2}))
\begin{equation}
  h(b)
  \leq \int_0^{\upi/2} (\frac{b}{\lamD} + 1) \,
                  \rme^{- b / \lamD} \rmd \theta
  = \frac{\upi}{2} \left( \frac{b}{\lamD} +1 \right) \, \rme^{-b / \lamD}
\end{equation}
implies (setting $b = \lamD \beta$ and $\beta = s-1$)
\begin{eqnarray}
  I_{12}
  & \leq &
  \frac{\upi^2}4  \int_{\lamD}^\infty
                        (\frac{b}{\lamD} +1)^2 \, \rme^{- 2 b / \lamD}  \,
                        b^{-1} \rmd b
  \nonumber \\
  & < &
  \frac{\upi^2}4 \int_1^\infty
                        (\beta +1)^2 \rme^{- 2 \beta}
                        \rmd \beta
  =
  \frac{\upi^2 \rme^2}{4}  \int_2^\infty   s^2  \rme^{- 2 s} \rmd s
  = A =
  \frac{13 \upi^2}{16 \rme^{2}}
  = 1.085\dots
\label{I12est}
\end{eqnarray}

On the other hand, the derivative of $h$ reads
\begin{eqnarray}
  h'(b)
  & = &
   \int_0^{\upi/2} \left[\frac{1}{\lamD}
       - \frac{\cos \theta +  b / \lamD} {\lamD \cos \theta}\right]
            \exp [- \frac{b}{\lamD \cos \theta}] \ \rmd \theta
  \nonumber \\
  & = &
   -  \frac{1} {\lamD} \int_0^{\upi/2}
      \frac{b}{ \lamD \cos \theta}
            \exp [- \frac{b}{\lamD \cos \theta}] \ \rmd \theta
\end{eqnarray}
where the divergent first factor is tamed by the exponential vanishing of the second factor for $\theta \to \upi/2$.
Now, for $0 \leq x < \infty$, the function $u(x) = x \rme^{-x}$ is maximum at $x=1$, so that
\begin{equation}
  h'(b) \, \lamD
  \geq
   -  \int_0^{\upi/2}
      \frac{1}{ \rme} \ \rmd \theta
  =
  -  c
\end{equation}
with $c = \upi / (2 \rme) = 0.577\ldots$,
and we estimate $I_{11}$ using
$1 \geq h(b) \geq 1 - c b / \lamD$,
\begin{eqnarray}
  - I_{11}(\lambda)
  & = &
  \int_\lambda^{\lamD}
    \frac {1 - h^2(b)}{b} \ \rmd b
  \leq
  \int_\lambda^{\lamD}
    \left( 2 \frac{c}{ \lamD} - \frac {c^2 b}{\lamD^2} \right) \ \rmd b
  \leq
  \left[ 2 \frac{c b}{\lamD} - \frac{c^2 b^2}{2  \lamD^2}
    \right]_\lambda^{\lamD}
  \nonumber \\
  & \leq &
  B =
 2 c - \frac{c^2}{2} = 0.988\dots
\label{I11est}
\end{eqnarray}
Eqs (\ref{I1sum}), (\ref{I12est}) and (\ref{I11est}) show that  $I_1(0)$ is bounded by
$- B \leq I_1(0) \leq A$. Numerically, we find
$I_1(0) = - 0.38\dots$

Finally, the inequality $h^2 \leq 1$ implies that $I_1(\lma)$ is an increasing function of $\lma$ between $0$ and $\lamD$.
Of course, the limit $\lma \to 0$ cannot be taken for the full integral $I$ of Eq.\ (\ref{I0I1}), for this limit is in the close collision regime.

Returning to Eq.\ (\ref{A:estimD}), we also note that the integral over $\bsv_j$, using spherical coordinates for $\bsv_j - \bsv_l$,
eliminates the denominator $\Delta v_{lj}$.
The dependence of $\lma$ on $\Delta v_{lj}$ implies that the trace $T_D$ of the diffusion tensor will finally depend on the temperature,
leading to the dominant contribution $\ln (\lamD / \lambda_{\mathrm{cma}})$ resulting from Eq.\ (\ref{logC}).
The overall result of taking the actual function $h$ into account
thus amounts to adding a finite constant $C'$ to the Coulomb logarithm.



\bibliographystyle{jpp}

\bibliography{EED_collisions_v13}

\end{document}